\documentclass[preprint,aps]{revtex4}

\usepackage{graphicx}
\usepackage{dcolumn}
\usepackage{bm}

\begin{document}

\title{Time-dependent Displaced and Squeezed Number States}

\author{Sang Pyo Kim}\email{sangkim@kunsan.ac.kr}

\affiliation{Department of Physics, Kunsan National University,
Kunsan 573-701, Korea}

\date{\today}
\begin{abstract}
We generalize the wave functions of the displaced and squeezed
number states, found by Nieto, to a time-dependent harmonic
oscillator with variable mass and frequency. These time-dependent
displaced and squeezed number states are obtained by first
squeezing and then displacing the exact number states and are
exact solutions of the Schr\"{o}dinger equation. Further, these
wave functions are the time-dependent squeezed harmonic-oscillator
wave functions centered at classical
trajectories. 
\end{abstract}

\maketitle

Recently, Nieto found the wave functions of the displaced and
squeezed number states for a static oscillator \cite{nieto}. These
states are an extension of the displaced (coherent) and squeezed
states of the vacuum state. The coherent states, actually
discovered by Sch\"{o}dinger \cite{sch}, have been developed as a
useful concept and tool in quantum optics \cite{coh}. A general
class of Gaussian states beyond the vacuum state was first
considered in Ref. 4, and these squeezed states were defined
further in terms of the squeeze operator \cite{sto}. The displaced
and squeezed states were also introduced \cite{roy}, and the
displaced number states, as well as the squeezed number states,
were found \cite{knight}. The squeezed number states show the
super-Poissonicity of photon statistics.

On the other hand, Lewis and Riesenfeld \cite{lew} introduced the
invariant method to find the Fock space of exact states of
time-dependent oscillators (for a review and references, see Refs.
9 and 10). Based on the invariant method, the coherent states of
time-dependent oscillators were found \cite{dod,har,kim-lee}, and
the squeezed states were studied \cite{ali,kim1,kim2}. Through the
use of the evolution in terms of su(1,1) algebra, the displaced
and squeezed number states were found for a general driven
time-dependent oscillator \cite{lo}.

The purpose of this paper is to find the displaced and squeezed
number states and their wave functions for a time-dependent
oscillator with variable mass and frequency. A time-dependent
oscillator also has time-dependent annihilation and creation
operators, the invariant operators satisfying the quantum
Liouville-von Neumann equation, whose number states are the exact
solutions of the time-dependent Schr\"{o}dinger equation
\cite{dod,kim1,kim2,kim-page,ksk,kim3}. We find the invariant
displacement and squeeze operators that map the exact number
states into the displaced and squeezed number states. For a static
oscillator, we compare these wave functions with those of Ref. 1.

We consider a time-dependent oscillator with variable mass and
frequency and with a Hamiltonian given by
\begin{equation}
\hat{H} (t) = \frac{1}{2m (t)} \hat{p}^2 + \frac{1}{2} m(t)
\omega^2 (t) \hat{x}^2, \label{osc}
\end{equation}
and use the invariant operator method to find the Fock space of
exact number states. Two linear invariant operators are introduced
\cite{dod,kim1,kim2,kim-page}:
\begin{eqnarray}
\hat{a} (t)  &=& \frac{i}{\sqrt{\hbar}} [u^*(t) \hat{p}
- m(t) \dot{u}^*(t)  \hat{x} ], \nonumber\\
\hat{a}^{\dagger} (t) &=& - \frac{i}{\sqrt{\hbar}} [u (t) \hat{p}
- m(t) \dot{u} (t) \hat{x}], \label{inv op}
\end{eqnarray}
where $u$ is a complex solution to the classical equation of
motion,
\begin{equation}
\ddot{u} (t) + \frac{\dot{m}(t)}{m (t)} \dot{u} (t) + \omega^2 (t)
u (t) = 0. \label{cl eq}
\end{equation}
With the Wronskian condition,
\begin{equation}
m (t) [u (t) \dot{u}^* (t) - u^* (t) \dot{u}(t)] = i, \label{wron}
\end{equation}
imposed, the operators in Eq. (\ref{inv op}) satisfy the usual
commutation relation at equal times
\begin{equation}
[\hat{a} (t), \hat{a}^{\dagger} (t) ] = 1,
\end{equation}
and play the roles of time-dependent annihilation and creation
operators. As the Hamiltonian in Eq. (\ref{osc}) has the
representation
\begin{equation}
\hat{H} (t) = \frac{\hbar m}{2} \bigl[( \dot{u}^* \dot{u} +
\omega^2 u^* u ) (\hat{a}^{\dagger} \hat{a} + \hat{a}
\hat{a}^{\dagger}) + (\dot{u}^{*2} + \omega^2 u^{*2})
\hat{a}^{\dagger 2} + (\dot{u}^2 + \omega^2 u^2) \hat{a}^2 \bigr],
\end{equation}
$\hat{a}(t)$ and $\hat{a}^{\dagger} (t)$ do not necessarily
diagonalize the Hamiltonian. The condition for diagonalization of
the Hamiltonian is given by the equation
\begin{equation}
\dot{u}^2 (t) + \omega^2 u^2 (t) = 0. \label{diag}
\end{equation}
Equation (\ref{diag}) is exactly satisfied by $u (t) = e^{- i
\omega_0 t}/\sqrt{2 m_0 \omega_0}$ for a static oscillator with
constant $m_0$ and $\omega_0$ and is approximately satisfied by
the adiabatic (WKB) solution, $u (t) = e^{- i \int \omega (t)
}/\sqrt{2 m \omega(t)}$, for slowly varying mass and frequency.
Except for these cases, the number operator $\hat{N} (t) =
\hat{a}^{\dagger} (t) \hat{a} (t)$ does not commute with the
Hamiltonian $\hat{H} (t)$. However, the number operator $\hat{N}
(t)$, which is another invariant operator, still provides the
exact wave functions of number states for the time-dependent
Schr\"{o}dinger equation \cite{kim-page}:
\begin{eqnarray}
\Psi_n (x, u(t)) = \Biggl(\frac{1}{2^{n} n! \sqrt{2 \pi \hbar}
\rho} \Biggr)^{1/2} e^{- i \Theta (n + 1/2)} H_n
\Biggl(\frac{x}{\sqrt{2 \hbar} \rho} \Biggr) \exp \Biggl[\frac{i m
\dot{u}^*}{2 \hbar u^*} x^2\Biggr], \label{wave}
\end{eqnarray}
where $u (t) = \rho (t) e^{- i \Theta (t)}$.

As Eq. (\ref{cl eq}) is linear, the most general complex solution
can take the form
\begin{equation}
u_{\nu} (t) = \mu u_0 (t) + \nu u_0^* (t), \label{most sol}
\end{equation}
where
\begin{eqnarray}
\mu = \cosh r, \quad \nu = e^{- i \phi} \sinh r, \label{sq par}
\end{eqnarray}
and $u_0 (t)$ is some preferred complex solution that satisfies
Eq. (\ref{wron}). The preferred solution may be chosen based on,
for instance, the minimum uncertainty or energy expectation value
\cite{kim1,kim2}. Here, we have fixed the overall phase of Eq.
(\ref{sq par}) since the wave functions in Eq. (\ref{wave}) do not
depend on it. In fact, $\nu$ (or the two real constants, $r$ and
$\phi$) is the squeezing parameter of the Bogoliubov
transformation
\begin{eqnarray}
\hat{a}_{\nu} (t) &=& (\cosh r) \hat{a}_0 (t) - (e^{ i \phi} \sinh
r) \hat{a}^{\dagger}_0 (t),
\nonumber\\
\hat{a}^{\dagger}_{\nu} (t) &=& (\cosh r) \hat{a}^{\dagger}_0 (t)
- (e^{- i \phi} \sinh r) \hat{a}_0 (t), \label{bog tran}
\end{eqnarray}
where $\hat{a}_{\nu}(t), \hat{a}^{\dagger}_{\nu}(t)$, and
$\hat{a}_0(t), \hat{a}^{\dagger}_0(t)$ are the operators obtained
by substituting $u_{\nu}(t)$ and $u_0(t)$, respectively, into Eq.
(\ref{inv op}). Note that the subscript 0 in this paper denotes
the zero squeezing parameter instead of a static oscillator with
constant frequency and mass. In terms of the squeeze operator
\cite{sto}
\begin{equation}
\hat{S} (\nu, t) = \exp \Biggl[ \frac{1}{2} (\nu e^{i \pi})
\hat{a}^{\dagger 2}_0 (t) - \frac{1}{2} (\nu e^{ i \pi})^*
 \hat{a}^2_0(t) \Biggr], \label{sq op}
\end{equation}
the Bogoliubov transformation is written as
\begin{eqnarray}
\hat{a}_{\nu} (t) &=& \hat{S}(\nu, t) ~\hat{a}_0 (t)~
\hat{S}^{\dagger}(\nu, t), \nonumber\\
\hat{a}_{\nu}^{\dagger} (t) &=& \hat{S}(\nu, t) ~
\hat{a}_0^{\dagger} (t) ~ \hat{S}^{\dagger}(\nu, t). \label{un
tran}
\end{eqnarray}

The number operators $ \hat{N}_0 (t) = \hat{a}_0^{\dagger} (t)
\hat{a}_0 (t)$ and $ \hat{N}_{\nu} (t) = \hat{a}^{\dagger}_{\nu}
(t) \hat{a}_{\nu} (t)$ do not, in general, commute with each
other,
\begin{equation}
[ \hat{N}_{\nu} (t), \hat{N}_0 (t) ] = \sinh 2r [e^{i \phi}
\hat{a}_0^{\dagger 2}(t) - e^{- i \phi} \hat{a}_0^2(t)],
\end{equation}
though they commute only for the trivial case of zero squeezing
$(r = 0)$. Nevertheless, they, being invariant operators, lead to
the exact number states for the oscillator in Eq. (\ref{osc}),
\begin{eqnarray}
\hat{N}_{0} (t) \vert n, t \rangle &=& n \vert n,  t \rangle, \nonumber\\
\hat{N}_{\nu} (t) \vert \nu, n, t \rangle &=& n \vert \nu, n, t
\rangle, \label{num st}
\end{eqnarray}
and to their wave functions
\begin{eqnarray}
\langle x \vert n, t \rangle = \Psi_n (x, u_0(t)), \nonumber\\
\langle x \vert \nu, n, t \rangle = \Psi_n (x, u_{\nu}(t)).
\label{num wave}
\end{eqnarray}
Here $\Psi_n (x, u_{0}(t))$ and $\Psi_n (x, u_{\nu}(t))$ are the
wave functions in Eq. (\ref{wave}) obtained by using $u_0(t)$ and
$u_{\nu}(t)$, respectively. Due to the unitary transformation in
Eq. (\ref{un tran}), the $\nu$-dependent number state is a
squeezed state,
\begin{equation}
\vert \nu, n, t \rangle = \hat{S} (\nu, t) \vert n, t \rangle.
\label{sq nst}
\end{equation}

The wave function $\Psi_n (x, u_{\nu} (t))$ of Eq. (\ref{sq nst})
is an exact solution for the Hamiltonian in Eq. (\ref{osc}), which
belongs to the wave functions in Eq. (\ref{wave}). In an other
way, we can directly show that any invariant operator $\hat{O}(t)$
maps an exact state $\vert \Psi_0 (t) \rangle $ to another state
$\hat{O}(t)\vert \Psi_0 (t) \rangle$. In fact, the quantum
Liouville-von Neumann equation,
\begin{equation}
i \hbar \frac{\partial}{\partial t} \hat{O} (t) + [\hat{O}(t),
\hat{H} (t)] = 0,
\end{equation}
leads to the time-dependent Schr\"{o}dinger equation
\begin{equation}
i \hbar \frac{\partial}{\partial t} \Bigl(\hat{O}(t)\vert \Psi_0
(t) \rangle \Bigr) = \hat{H} (t) \Big(\hat{O}(t)\vert \Psi_0 (t)
\rangle \Bigr) \label{map eq}
\end{equation}
whenever $\vert \Psi_0 (t) \rangle$ is a solution to the
time-dependent Schr\"{o}dinger equation. Note that the squeeze
operator in Eq. (\ref{sq op}) is an invariant operator since it is
a functional of the invariant operators $\hat{a}_0 (t)$ and
$\hat{a}_0^{\dagger} (t)$. Thus, $\hat{S} (\nu, t)$ generates the
exact state $\vert \nu, n, t \rangle$ out of the number state
$\vert n, t \rangle$, which is already an exact one.

As for the case of a time-independent oscillator \cite{sto}, the
displaced (coherent) state is an eigenstate of the time-dependent
annihilation operator \cite{har,kim2},
\begin{equation}
\hat{a}_{\nu} (t) \vert \alpha, \nu, t \rangle = \alpha \vert
\alpha, \nu, t \rangle,
\end{equation}
for any complex constant $\alpha$, and has the representation
\begin{equation}
\vert \alpha, \nu, t \rangle = e^{- |\alpha|^2/2} \sum_{n =
0}^{\infty} \frac{\alpha^n}{\sqrt{n!}} \vert \nu, n, t \rangle.
\label{num rep}
\end{equation}
The time-dependent displaced state in Eq. (\ref{num rep}) is an
exact quantum state of the Schr\"{o}dinger equation for the
Hamiltonian in Eq. (\ref{osc}) since each $\vert \nu, n, t
\rangle$ is a solution to the Schr\"{o}dinger equation. In another
formulation, we introduce the time-dependent displacement operator
\begin{equation}
\hat{D} (\alpha, t) = e^{\alpha \hat{a}^{\dagger}_{\nu} (t) -
\alpha^* \hat{a}_{\nu} (t)}, \label{dis op}
\end{equation}
which is a functional of the invariant operators $\hat{a}_0 (t)$
and $\hat{a}_0^{\dagger} (t)$ and, thus, is an invariant operator.
If the operator in Eq. (\ref{dis op}) operates on the squeezed
number state in Eq. (\ref{sq nst}), we obtain the displaced and
squeezed number state
\begin{equation}
\vert \alpha, \nu, n, t \rangle = \hat{D} (\alpha, t) \vert \nu,
n, t \rangle  =  \hat{D} (\alpha, t) \hat{S} (\nu, t) \vert n, t
\rangle. \label{coh nst}
\end{equation}
Using the relations
\begin{eqnarray}
\hat{D} (\alpha, t)~ \hat{a}_{\nu} (t) ~\hat{D}^{\dagger}(\alpha,
t) = \hat{a}_{\nu} (t) - \alpha, \nonumber\\
\hat{D} (\alpha, t) \hat{a}^{\dagger}_{\nu} (t)
\hat{D}^{\dagger}(\alpha, t) = \hat{a}_{\nu}^{\dagger} (t) -
\alpha^*, \label{dis rel}
\end{eqnarray}
we get another representation
\begin{eqnarray}
\vert \alpha, \nu, n, t \rangle &=& \frac{1}{\sqrt{n!}}
[\hat{a}_{\nu} (t) - \alpha^*]^n \vert \alpha, \nu, t
\rangle \nonumber\\
&=& \frac{e^{- | \alpha |^2}}{\sqrt{n!}} \sum_{k = 0}^{\infty}
\frac{\alpha^k}{\sqrt{k!}} [\hat{a}_{\nu} (t) - \alpha^*]^n \vert
\nu, k, t \rangle. \label{num rep2}
\end{eqnarray}
Being a sum of exact number states, the displaced and squeezed
number state in Eq. (\ref{coh nst}) or (\ref{num rep2}) is another
exact quantum state of the time-dependent Schr\"{o}dinger
equation. In an other way, $\hat{D} (\alpha, t) \hat{S} (\nu, t)$,
a product of invariant operators, is still another invariant
operator and, thus, maps the exact number state $\vert n, t
\rangle$ into the displaced and squeezed number state $\vert
\alpha, \nu, n, t \rangle$, which, according to Eq. (\ref{map
eq}), is an exact quantum state of the time-dependent
Schr\"{o}dinger equation for the Hamiltonian in Eq. (\ref{osc}).

The displaced and squeezed number state has the expectation values
\begin{eqnarray}
x_c (t) &=& \langle \alpha, \nu, n, t \vert \hat{x} \vert \alpha,
\nu, n, t \rangle = \sqrt{\hbar} [\alpha u_{\nu}
(t) + \alpha^* u^*_{\nu} (t)], \nonumber\\
p_c (t) &=& \langle \alpha, \nu, n, t \vert \hat{p} \vert \alpha,
\nu, n, t \rangle = \sqrt{\hbar} m(t) [\alpha \dot{u}_{\nu} (t) +
\alpha^* \dot{u}^*_{\nu} (t)], \label{coh ex}
\end{eqnarray}
and
\begin{eqnarray}
\langle \alpha, \nu, n, t \vert \hat{x}^2 \vert \alpha, \nu, n, t
\rangle &=&
\hbar u_{\nu}^* u_{\nu} (2n+1) + x_c^2,\nonumber\\
\langle \alpha, \nu, n, t \vert \hat{p}^2 \vert \alpha, \nu, n, t
\rangle &=& \hbar m^2 \dot{u}_{\nu}^* \dot{u}_{\nu} (2n+1) +
p_c^2.
\end{eqnarray}
Note that $p_c (t) = m(t) \dot{x}_c(t)$, and that $x_c(t)$ obeys
the classical equation of motion in Eq. (\ref{cl eq}). Inverting
Eq. (\ref{coh ex}) for
\begin{eqnarray}
\alpha &=& \frac{i}{\sqrt{\hbar}} [ u^*_{\nu} (t) p_c (t) - m(t)
\dot{u}^*_{\nu}
(t) x_c (t)], \nonumber\\
\alpha^* &=& - \frac{i}{\sqrt{\hbar}} [ u_{\nu} (t) p_c (t) - m(t)
\dot{u}_{\nu} (t) x_c (t)],
\end{eqnarray}
and using Eq. (\ref{dis rel}), we finally obtain the wave function
for the displaced and squeezed number state:
\begin{eqnarray}
\Psi_n (x, x_c, p_c, u_{\nu}(t)) = \Biggl(\frac{1}{2^{n} n!
\sqrt{2 \pi \hbar} \rho_{\nu}} \Biggr)^{1/2} e^{- i \Theta_{\nu}
(n + 1/2)} e^{i p_c x/ \hbar}
\nonumber\\
\times H_n \Biggl(\frac{x - x_c}{\sqrt{2 \hbar} \rho_{\nu}}
\Biggr) \exp \Biggl[\frac{i m \dot{u}_{\nu}^*}{2 \hbar u_{\nu}^*}
(x - x_c)^2\Biggr], \label{coh sq wave}
\end{eqnarray}
where $u_{\nu} (t) = \rho_{\nu} (t) e^{- i \Theta_{\nu}(t)}$.
Thus, the wave functions in Eq. (\ref{coh sq wave}) are, up to a
phase factor $e^{i p_c x/ \hbar}$, the wave functions in Eq.
(\ref{wave}) with the center shifted to the classical trajectory
$x_c (t)$.

Finally, we compare the wave functions in Eq. (\ref{coh sq wave})
of the displaced and squeezed number states for a time-dependent
oscillator with those of Ref. 1 for a static oscillator. The
static oscillator has a constant mass $m_0$  and a frequency
$\omega_0$. We may then choose
\begin{equation}
u_0 (t) = \frac{1}{\sqrt{2 m_0 \omega_0}} e^{- i \omega_0 t},
\label{pref sol}
\end{equation}
as the preferred solution to Eq. (\ref{cl eq}). If $u_0(t)$ is
inserted into Eq. (\ref{wave}), it leads to the
harmonic-oscillator wave functions. Now, the displaced state with
the most general solution, Eq. (\ref{most sol}), has the
expectation value
\begin{eqnarray}
x_c (t) &=& \sqrt{\frac{\hbar}{2 m_0 \omega_0}} \Bigl[(\alpha
\cosh r - \alpha^* e^{i \phi} \sinh r) e^{- i \omega_0 t} +
(\alpha^* \cosh r - \alpha e^{- i \phi} \sinh r) e^{i \omega_0 t}
\Bigr] \nonumber\\& \equiv & x_0 \cos \omega_0 t + \frac{p_0}{m_0
\omega_0} \sin \omega_0 t,
\end{eqnarray}
and $p_c (t) = m_0 \dot{x}_c (t)$. The classical trajectory has
two integration constants: $x_c (0) = x_0$ and $p_c (0) = p_0$.
Then, the wave function in Eq. (\ref{coh sq wave}) becomes
\begin{eqnarray}
\Psi_n (\nu, x_c, p_c, x, t) = \frac{1}{\sqrt{2^n n!}} \Biggl(
\frac{A_{\nu}}{\sqrt{\pi}} \Biggr)^{1/2} e^{- i \Theta_{\nu} (n +
1/2)} e^{i p_c x/ \hbar} \nonumber\\ \times H_n (A_{\nu} (x -
x_c)) e^{- B_{\nu} (x - x_c)^2}, \label{gen wave}
\end{eqnarray}
where
\begin{eqnarray}
A_{\nu} (t) &=& \frac{1}{\sqrt{2 \hbar u_{\nu}^* u_{\nu}}} =
\sqrt{\frac{m_0 \omega_0 }{\hbar}} \frac{1}{[\cosh 2r + \sinh 2r
\cos (2 \omega_0 t - \phi)]^{1/2}},
\nonumber\\
B_{\nu} (t) &=& \frac{i m \dot{u}_{\nu}^*}{2 \hbar u_{\nu}^*} =
\frac{m_0 \omega_0}{2 \hbar} \Biggl[\frac{\cosh r e^{i \omega_0 t}
- e^{i \phi} \sinh r e^{- i \omega_0 t} }{\cosh r e^{i \omega_0 t}
+
e^{i \phi} \sinh r e^{- i \omega_0 t}} \Biggr], \nonumber\\
e^{ - i \Theta_{\nu} (t)} &=& \Biggl[ \frac{\cosh r e^{- i
\omega_0 t} +  e^{- i \phi} \sinh r e^{i \omega_0 t}}{\cosh r e^{
i \omega_0 t} +  e^{i \phi} \sinh r e^{- i \omega_0 t}}
\Biggr]^{1/2}. \label{coeff}
\end{eqnarray}
We can show, for the convention, $m_0 = \omega_0 = \hbar = 1$,
that
\begin{eqnarray}
A_{\nu} (0) = \frac{1}{{\cal F}_4}, \quad B_{\nu} (0) =
\frac{{\cal F}_2}{2}, \quad e^{- i \Theta_{\nu} (0)} = {\cal
F}_3^{1/2},
\end{eqnarray}
where ${\cal F}$'s in Eqs. (21)-(24) of Ref. 1 are given by
\begin{eqnarray}
{\cal F}_2 &=& \frac{1 - i \sin \phi \sinh r (\cosh r + e^{i \phi}
\sinh r)}{(\cosh r + \cos \phi \sinh r) (\cosh r + e^{i \phi}
\sinh
r)}, \nonumber\\
{\cal F}_3 &=& \frac{\cosh r + e^{- i \phi} \sinh r}{\cosh r +
e^{i \phi} \sinh r}, \nonumber\\
{\cal F}_4 &=& (\cosh^2 r + \sinh^2 r + 2 \cos \phi \cosh r \sinh
r)^{1/2}.
\end{eqnarray}
Thus, the wave functions in Eq. (20) of Ref. 1 for the static
oscillator are special cases of the wave functions in Eq.
(\ref{gen wave}) at $t = 0$. Similarly, we show that
\begin{eqnarray}
A_{\nu} (t) = \frac{1}{{\cal F}_4 B A^{1/2}}, \quad B_{\nu} (t) =
\frac{{\cal F}_2 \cos t + i \sin t}{2(\cos t + i {\cal F}_2 \sin
t)}, \quad e^{- i \Theta_{\nu} (t)} = ({\cal F}_3 A)^{1/2},
\end{eqnarray}
where $A$ in Eq. (46) and $B$ in Eq. (47) of Ref. 1 are given by
\begin{eqnarray}
A &=& \frac{{\cal F}_4^2 (\cos t + i {\cal F}_2 \sin t) - 2 i \sin
t}{{\cal F}_4^2 (\cos t + i {\cal F}_2 \sin t)}, \nonumber\\
B &=& \cos t + i {\cal F}_2 \sin t.
\end{eqnarray}
Hence, the wave functions in Eq. (\ref{gen wave}) of the
time-dependent Schr\"{o}dinger equation are the same as Eq. (45)
of Ref. 1 for the time-evolved displaced and squeezed states.

In summary, using the invariant method, we obtained the displaced
(coherent) and squeezed number states and their wave functions for
a time-dependent oscillator. These states for the time-dependent
oscillator are an extension of the wave functions found by Nieto
for a static oscillator. The wave functions for the displaced and
squeezed number states are found to be the squeezed
harmonic-oscillator wave functions centered around a classical
trajectory.

\acknowledgements

This work was supported by the Korea Research Foundation under
grant No. KRF-2002-041-C00053.

\end{document}